\documentclass[a4paper,12pt]{article}
\pdfoutput=1
\usepackage{jheppub}
\usepackage{tabularx} 
\usepackage{amsmath,amssymb}
\usepackage{graphicx} 
\graphicspath{ {./figures/} }
\usepackage{float}
\usepackage{tensor}
\usepackage{slashed}
\usepackage[compat=1.0.0]{tikz-feynman}

\begin{document}

\begin{flushright}
TU-1320,
KEK-QUP-2026-0018
\end{flushright}

\title{Gravitational Dark Matter Production in Einstein-Cartan Theory}

\author[a]{Kamil Mudruňka,}
\author[a,b]{Kazunori Nakayama}

\affiliation[a]{Department of Physics, Tohoku University, Sendai, Miyagi 980-8578, Japan}
\affiliation[b]{International Center for Quantum-field Measurement Systems for Studies of the Universe and Particles (QUP), KEK, Tsukuba, Ibaraki 305-0801, Japan}

\emailAdd{mudrunka.kamil.q5@dc.tohoku.ac.jp}
\emailAdd{kazunori.nakayama.d3@tohoku.ac.jp}

\abstract{Introducing torsion into General Relativity leads to new four-point interactions for fermions. These are universal and must also affect any fermion dark matter.
We improve existing calculations of dark matter production through the Einstein-Cartan theory four-fermion interaction by also considering the s-channel graviton exchange as another contributing process. The effect on both the total cross section as well as the momentum-dependent Boltzmann equation collision term is calculated and discussed. We also introduce the Holst term, which generally modifies the four-fermion interaction coupling constant as well as introduces two new four-fermion interactions, and discuss and compute its effect on the total produced dark matter abundance.}

\maketitle

\renewcommand{\thefootnote}{\fnsymbol{footnote}}
\renewcommand{\thefootnote}{\arabic{footnote}}
\setcounter{footnote}{0}

\section{Introduction}

The nature of dark matter remains an unsolved mystery to this day. While many existing and proposed experiments are designed around directly detecting dark matter particles through various proposed interactions, the possibility that dark matter interacts only through gravity remains. Despite the relative weakness of gravity compared to the other forces of nature, it is possible to achieve plausible dark matter production this way: the gravitational dark matter production during reheating (or the inflaton annihilation into the dark matter pair)~\cite{Ema:2015dka,Ema:2016hlw,Ema:2018ucl,Ema:2019yrd,Mambrini:2021zpp,Clery:2021bwz}, and thermal gravitational dark matter production~\cite{Garny:2015sjg,Tang:2016vch,Tang:2017hvq,Garny:2017kha}.
It has been also shown that the contribution from gravitational production can drastically affect the abundance of Higgs-portal scalar dark matter~\cite{Chianese:2020yjo,Aoki:2022dzd,Izumine:2024hyp}.

On the other hand, the nature of gravity should be identified in order to precisely evaluate the gravitational particle production rate.
General Relativity (GR) is built around the symmetric Levi-Civita connection. A natural generalization of the theory is to relax this assumption and treat the antisymmetric part of the connection, called torsion, as a new degree of freedom. 
This leads to the Einstein-Cartan (EC) theory, which links torsion to spin similarly to how energy-momentum tensor is linked to curvature. In the simplest version of EC, torsion becomes an algebraic constraint which can be integrated out to get ordinary GR with a new contact 4 fermion interaction.\footnote{
    Such fermion 4-point interactions also naturally appear in supergravity~\cite{Wess:1992cp}.
} 
While negligible at low energies, the interaction overtakes gravity at high energies. 
For example, EC theory may prevent the Big Bang singularity in certain models and replaces it with a bounce. Similarly, it also has the potential to prevent formation of black hole singularities~\cite{RevModPhys.48.393, Hehl:1971qi, Trautman:2006fp, Hashemi:2014nra, Poplawski:2011jz, Magueijo:2012ug}.

The fermion 4-point interactions in the EC theory may also affect the dark matter production in the early universe for fermionic dark matter.
In Ref.~\cite{Shaposhnikov:2020aen}, the dark matter abundance has been estimated, assuming that the dark matter particles are produced from the scattering of thermal bath through the 4-point interaction.
The purpose of this paper is to revisit the dark matter production in the EC theory. See also Ref.~\cite{Rigouzzo:2023sbb} for related issues.
Note that, as shown in Refs.~\cite{Garny:2015sjg,Tang:2016vch,Tang:2017hvq,Garny:2017kha}, the graviton exchange processes also contribute to the dark matter production, which is naively the same order as that from the 4-point interaction in the EC theory.
We fully take into account the gravitational production processes, and study possible interference between the purely gravitational portal and the EC interaction portal. We also generalize the results to other possible torsion-induced dimension 6 operators. 

This paper is organized as follows. 
In Sec.~\ref{sec:EC}, we review the basics of EC theory and derive its four-fermion interaction.
In Sec.~\ref{sec:EConly}, we calculate the dark matter production rate taking account of both the s-channel graviton exchange and the EC 4-point coupling, and show that the interference between the s-channel graviton exchange and EC interaction portal has vanishing contribution to the total cross section. 
In Sec.~\ref{sec:holst}, we introduce the other possible dimension 6 four-fermion interactions by adding the Holst term and compute their interference with both the s-channel graviton exchange and among themselves. 
Sec.~\ref{sec:conc} is devoted to conclusions and discussion.
Appendix~\ref{sec:massive} lists the matrix elements for the calculations done in this paper in the general case of massive dark matter and massless Standard Model (SM) fermions. 
In Appendix~\ref{sec:ang}, we study the angular distribution of the produced dark matter and show how the quadrupole anisotropy can be removed by tuning the EC interaction coupling constant. 
In Appendix~\ref{sec:integral}, we present technical details of the phase space integral.

All the matrix elements presented in the paper are spin summed unless it is stated otherwise. We consider the dark matter fermions to be Dirac spinors. Throughout the paper we use $\kappa^2=32\pi G=4/M_{\rm pl}^2$. We work in the mostly minus $+,-,-,-$ metric convention. 

\section{Einstein-Cartan theory} \label{sec:EC}

Einstein-Cartan (EC) theory is a natural extension of GR which relaxes the torsion free connection assumption and links torsion to spin. The torsion tensor is defined as the antisymmetric part of the connection
\begin{equation}
    \tensor{T}{^\alpha_\mu_\nu}=\tensor{\Gamma}{^\alpha_\mu_\nu}-\tensor{\Gamma}{^\alpha_\nu_\mu}.
\end{equation}
A general connection might also contain a nonmetricity term resulting in $\nabla_{\alpha}g_{\mu\nu}\neq0$, but we will only consider metric compatible connections here. For practical purposes torsion can be equivalently described by the contorsion tensor $K_{\mu\nu\alpha}$, which we define as
\begin{equation}
    K_{\mu\nu\alpha}=\frac{1}{2}\left(T_{\alpha\mu\nu}-T_{\nu\alpha\mu}-T_{\mu\nu\alpha}\right).
\end{equation}
This allows us to write the connection components as
\begin{equation}
    \tensor{\Gamma}{^\alpha_\mu_\nu}=\tensor{\bar{\Gamma}}{^\alpha_\mu_\nu}+\tensor{K}{^\alpha_\nu_\mu},
\end{equation}
where $\tensor{\bar{\Gamma}}{^\alpha_\mu_\nu}$ are the Christoffel symbols or the Levi-Civita connection. One thing to notice is that the contorsion tensor is antisymmetric in its first two indices, while the torsion tensor is antisymmetric in the last two indices. We can decompose $K_{\mu\nu\alpha}$ into its trace, antisymmetric and remaining traceless parts \cite{Shapiro:2001rz}
\begin{equation}
    K_\nu=\tensor{K}{^\alpha_\nu_\alpha},~~~S_{\alpha\mu\nu}=K_{[\alpha\mu\nu]},~~~P_{\alpha\mu\nu}=K_{\alpha\mu\nu}-\frac{2}{3}K_{[\mu} g_{\alpha]\nu}-S_{\alpha\mu\nu}.
\end{equation}
Using the Hodge dual we can alternatively represent $S_{\alpha\mu\nu}$ through an axial vector $S^\beta$
\begin{equation}
    S_{\alpha\mu\nu}=\frac{\sqrt{-g}}{6}\epsilon_{\beta\alpha\mu\nu} S^\beta.
\end{equation}
The EC theory assumes the standard Einstein-Hilbert action for gravity
\begin{equation}
    S_{\rm EH}=-\int d^4 x\sqrt{-g}\frac{2}{\kappa^2}R,
\end{equation}
which in the presence of totally antisymmetric torsion becomes
\begin{equation}
\label{eq:ECaction}
    S_{\rm EH}=-\int d^4 x\sqrt{-g}\frac{2}{\kappa^2}\left(\bar{R}-K^{\mu\nu\alpha}K_{\mu\nu\alpha}\right),
\end{equation}
where $\bar{R}$ is the Ricci scalar built only out of the Levi-Civita connection. There is no kinetic term for torsion so the equations we obtain by taking variations with respect to $K$ are algebraic constraints. While torsion does not appear in the usual Lagrangians of scalar and gauge fields, it naturally couples to fermions. The Dirac equation can be written as
\begin{equation}
    (i \gamma^{\mu}\mathcal{D}_\mu-m)\psi=0,
\end{equation}
where $\mathcal{D}$ is the covariant derivative given by
\begin{equation}
    \mathcal{D}_{\mu}=\partial_\mu+\frac{1}{2}\omega_\mu^{ab}\sigma_{ab},~~~\sigma_{ab}=\frac{1}{4}\left[\gamma_a,\gamma_b\right],
\end{equation}
$\omega^{ab}_\mu$ is the 1-form connection corresponding to the vierbein, and the Dirac matrices with spacetime indices are constructed using the vierbein as $\gamma^\mu =e^\mu_a \gamma^a$. To see how torsion couples to fermions we write down the action
\begin{equation}
    S=\frac{1}{2}\int d^4x\sqrt{-g}\left(i\bar{\psi}\gamma_\alpha\mathcal{D}^\alpha\psi-i\mathcal{D}^\alpha\bar{\psi}\gamma_\alpha\psi-2m\bar{\psi}\psi\right).
\end{equation}
The connection contribution is 
\begin{equation}
    \frac{i}{4}\int d^4x\sqrt{-g}\left[\omega^{\mu\alpha\beta}\bar{\psi}\gamma_\mu \sigma_{\alpha\beta}\psi-\omega^{\mu\beta\alpha}\bar{\psi}\sigma_{\alpha\beta}\gamma_\mu\psi\right].
\end{equation}
Using the identity
\begin{equation}
    \frac{1}{2}\gamma^a\left[\gamma^b,\gamma^c\right]=\frac{1}{4}\lbrace\gamma^a,\left[\gamma^b,\gamma^c\right]\rbrace+\eta^{ab}\gamma^c-\eta^{ac}\gamma^b,
\end{equation}
where the first term on the right hand side is totally antisymmetric and the fact that the contorsion is antisymmetric in the first two indices, we obtain the action
\begin{equation}
    S_F=\frac{1}{2}\int d^4x\sqrt{-g}\left(i\bar{\psi}\gamma_\alpha\bar{\mathcal{D}}^\alpha\psi-i\bar{\mathcal{D}}^\alpha\bar{\psi}\gamma_\alpha\psi-2m\bar{\psi}\psi+\frac{i}{2}\bar{\psi}K_{\mu\alpha\beta}\gamma^{[\mu}\gamma^\alpha\gamma^{\beta]}\psi\right).
\end{equation}
We see that in EC theory only the totally antisymmetric part of torsion plays a nontrivial role so our restriction in \eqref{eq:ECaction} is justified. Finally integrating out torsion from the combined fermion and Einstein-Hilbert action gives us back pure GR with a new 4 fermion point interaction
\begin{align}
    &S=\int d^4x \sqrt{-g}\left[-\frac{2}{\kappa^2}\bar{R}+\frac{1}{2}\left(i\bar{\psi}\gamma_\alpha\bar{\mathcal{D}}^\alpha\psi-i\bar{\mathcal{D}}^\alpha\bar{\psi}\gamma_\alpha\psi-2m\bar{\psi}\psi\right)+\frac{3\kappa^2}{64}A_\mu A^\mu\right],
\end{align}
where
\begin{align}
    A_\mu=\bar{\psi}\gamma_5\gamma_\mu\psi.
\end{align}
If more fermion species exist, then there are also interactions between any two pairs
\begin{equation}
    S_{\rm int}=\frac{3\kappa^2}{64}\int d^4x\sqrt{-g}\sum_{ij}\left(\bar{\psi}_i\gamma_5\gamma_\mu\psi_i\right)\left(\bar{\psi}^j\gamma_5\gamma^\mu\psi^j\right).
    \label{eq:ECint}
\end{equation}
Note that in the above sum each pair appears twice in total, so the coupling becomes $3\kappa^2/32$ when considering two different species.

\section{Dark matter production in Einstein-Cartan theory} \label{sec:EConly}

\subsection{Fermion channel}

Thanks to the universality of the EC interaction \eqref{eq:ECint}, it will also necessarily affect dark matter production if it is of fermionic nature. We assume that the dark matter has no other gauge/Yukawa interactions besides gravity and EC interaction, and compute the matrix elements for purely gravitational dark matter production from SM fermion-antifermion pairs:
\begin{equation}
    \bar{\psi}_{\rm SM}+\psi_{\rm SM}\rightarrow\bar{\psi}_{\rm DM}+\psi_{\rm DM}.
\end{equation}
We will show that the resulting cross sections add up, i.e. there is no interference between the EC and gravitational paths. We present the results in the fully massless case for better readability. 
Our goal is to estimate the dark matter abundance created from high temperature plasma, and the matrix elements evaluated in the massless limit is enough for this purpose as far as the dark matter mass is smaller than the temperature. 
See Appendix~\ref{sec:massive} for the full formulas.

We take  the SM fermions to be chiral fields, while the dark matter is assumed to be Dirac fermion. The presented amplitudes are valid for both left and right handed species. First let us evaluate a contribution from the purely gravitational production path through the s-channel graviton exchange. It is described by one diagram
\begin{align}
\begin{tikzpicture}[baseline=(current bounding box.center)]
\begin{feynman}
\vertex (1) at (-1.5,0.7) {$\psi_{\rm SM}$};
\vertex[dot] (2) at (0,0.7) {};
\vertex (3) at (1.5,0.7) {$\psi_{\rm SM}$};
\vertex (4) at (-1.5,-0.7) {$\psi_{\rm DM}$};
\vertex[dot] (5) at (0,-0.7) {};
\vertex (6) at (1.5,-0.7) {$\psi_{\rm DM}$};
\diagram*{
  (2) -- [fermion] (1),
  (3) -- [fermion] (2),
  (2) -- [gluon,edge label=\(~h_{\mu\nu}\)] (5),
  (4) -- [fermion] (5),
  (5) -- [fermion] (6)
};
\end{feynman}
\end{tikzpicture}
\label{schannelgraviton}
\end{align}
Using the expression for the vertex derived from perturbing the Dirac Lagrangian,
\begin{align}
\begin{tikzpicture}[baseline=(current bounding box.center)]
\begin{feynman}
\vertex (1) at (-1.0,0.7) {$\psi$};
\vertex[dot] (2) at (0,0.7) {};
\vertex (3) at (1.0,0.7) {$\psi$};
\vertex (4) at (0,-0.7) {};
\diagram*{
  (1) -- [fermion] (2),
  (2) -- [fermion] (3),
  (2) -- [gluon] (4)
};
\end{feynman}
\end{tikzpicture}
=\frac{i}{8}\kappa\left[2\eta_{\mu\nu}\left(\slashed{p_1}-\slashed{p_2}-2m\right)-(p_1-p_2)_\mu \gamma_\nu -(p_1-p_2)_\nu \gamma_\mu\right],
\end{align}
and the graviton propagator in the De-Donder gauge,
\begin{equation}
\mu\nu~
    \vcenter{\hbox{\feynmandiagram [horizontal=i1 to i2] {
  i1 -- [gluon, momentum'=\(p\)] i2
};}}
~\alpha\beta\quad
    =\frac{i}{2}\frac{\eta^{\mu\alpha}\eta^{\nu\beta}+\eta^{\mu\beta}\eta^{\nu\alpha}-\eta^{\mu\nu}\eta^{\alpha\beta}}{p^2},
    \label{eq:gravprop}
\end{equation}
we find the squared matrix element to be~\cite{Garny:2017kha,Clery:2021bwz}
\begin{equation}
    \vert \mathcal M_{\rm G} \vert^2=\frac{\kappa^{4} \left(s^{4} + 10 s^{3} t + 42 s^{2} t^{2} + 64 s t^{3} + 32 t^{4}\right)}{256 s^{2}},
\end{equation}

Next let us consider the production through the EC interaction. We have only one diagram 
\begin{align}
\begin{tikzpicture}[baseline=(current bounding box.center)]
\begin{feynman}
\vertex[dot] (1) at (0,0) {};
\vertex (2) at (-1.,1.) {$\psi_{\rm SM}$};
\vertex (3) at (1.,1.) {$\psi_{\rm SM}$};
\vertex (4) at (-1.,-1.) {$\psi_{\rm DM}$};
\vertex (5) at (1.,-1.) {$\psi_{\rm DM}$};
\diagram*{
  (1) -- [fermion] (2),
  (3) -- [fermion] (1),
  (4) -- [fermion] (1),
  (1) -- [fermion] (5)
};
\end{feynman}
\end{tikzpicture}
\label{EC4point}
\end{align}
The corresponding interaction vertex is
\begin{equation}
    \frac{3i\kappa^2}{32}\left(\gamma_5\gamma^\mu\right)_{\rm SM}\left(\gamma_5\gamma_{\mu}\right)_{\rm DM}.
\end{equation}
The resulting spin summed matrix element is \cite{Shaposhnikov:2020aen}
\begin{equation}
    \vert \mathcal M_{\rm EC} \vert^2=\frac{9 \kappa^{4} \left[ (s+t)^2 + t^{2} \right]}{256}.
\end{equation}
The interference term between the two diagrams is
\begin{equation}
    2\Re\left(\mathcal M_{\rm EC} \mathcal M_{\rm G}^*\right)= \frac{3 \kappa^{4} \left(s^{2} + 6 s t + 6 t^{2}\right)}{128}.
\end{equation}
Going to the center-of-mass frame and writing 
\begin{equation}
    t=-\frac{s}{2}(1-\cos\theta),
\end{equation}
where $\theta$ is the angle between the incoming SM fermion momentum and outgoing dark matter momentum, we get 
\begin{equation}
    2 \Re\left(\mathcal M_{\rm EC} \mathcal M_{\rm G}^*\right)=-\frac{3\kappa^4 s^2}{256}\left(1-3\cos^2\theta\right)
\end{equation}
and so after integrating over all angles the interference term does not contribute to the total reaction rate
\begin{equation}
    \int d\Omega\,2 \Re(\mathcal M_{\rm EC}\mathcal M_{\rm G}^*)= 0.
\end{equation}
The amplitudes in terms of $\cos\theta$ are
\begin{equation}
     \vert \mathcal M_{\rm G} \vert^2=\frac{\kappa^4s^2}{512}\left(1-3\cos^2\theta+4\cos^4\theta\right),~~\vert \mathcal M_{\rm EC} \vert^2=\frac{9\kappa^4s^2}{512}\left(1+\cos^2\theta\right).
\end{equation}
Integrating the squared matrix elements over the full angle shows that the EC interaction dominates over the s-channel graviton exchange with the ratio between them being
\begin{equation}
    \frac{\int d\Omega |\mathcal M_{\rm EC}|^2}{\int d\Omega |\mathcal M_{\rm G}|^2}=15.
    \label{EC_ratio}
\end{equation}
Therefore, the s-channel graviton exchange contribution enhances the dark matter production rate by $6.7\,\%$ compared with the EC only result, if we only consider the SM fermion pair annihilation process.
However, in order to estimate the actual dark matter abundance, we must also take into account the s-channel graviton exchange with initial scalar/vector particles. The abundance will be estimated in Sec.~\ref{sec:abundance}.

\subsection{Scalar channel}

Dark matter also gets produced by a graviton mediated process of the Higgs field, which consists of four real scalar degrees of freedom.
For a real scalar particle $\phi$ the $\phi+\phi\rightarrow\bar{\psi}_{\rm DM}+\psi_{\rm DM}$ is described by the diagram
\begin{align}
\begin{tikzpicture}[baseline=(current bounding box.center)]
\begin{feynman}
\vertex (1) at (-1.5,0.7) {$\phi$};
\vertex[dot] (2) at (0,0.7) {};
\vertex (3) at (1.5,0.7) {$\phi$};
\vertex (4) at (-1.5,-0.7) {$\psi_{\rm DM}$};
\vertex[dot] (5) at (0,-0.7) {};
\vertex (6) at (1.5,-0.7) {$\psi_{\rm DM}$};
\diagram*{
  (2) -- [scalar] (1),
  (3) -- [scalar] (2),
  (2) -- [gluon,edge label=\(~h_{\mu\nu}\)] (5),
  (4) -- [fermion] (5),
  (5) -- [fermion] (6)
};
\end{feynman}
\end{tikzpicture}
\label{scalarschannelgraviton}
\end{align}
where the $\phi\phi h$ vertex is
\begin{align}
\begin{tikzpicture}[baseline=(current bounding box.center)]
\begin{feynman}
\vertex (1) at (-1.0,0.7) {$\phi$};
\vertex[dot] (2) at (0,0.7) {};
\vertex (3) at (1.0,0.7) {$\phi$};
\vertex (4) at (0,-0.7) {};
\diagram*{
  (1) -- [scalar] (2),
  (2) -- [scalar] (3),
  (2) -- [gluon] (4)
};
\end{feynman}
\end{tikzpicture}
=-\frac{i}{2}\kappa\left(p_{1\mu}p_{2\nu}+p_{2\mu}p_{1\nu}-\eta_{\mu\nu}p_1\cdot p_2\right),
\end{align}
Summing over the dark matter helicities, we get
\begin{equation}
    \vert \mathcal M_{\phi} \vert^2= \frac{\kappa^4}{32s^2}tu(t-u)^2
\end{equation}
In the center of mass frame this becomes
\begin{equation}
     \vert \mathcal M_{\phi} \vert^2=\frac{\kappa^4s^2}{128}\cos^2\theta(1-\cos^2\theta).
\end{equation}

\subsection{Gauge boson channel}

Finally the SM contains the strong and electroweak gauge bosons and they also contribute to the production. The relevant diagram is
\begin{align}
\begin{tikzpicture}[baseline=(current bounding box.center)]
\begin{feynman}
\vertex (1) at (-1.5,0.7) {$A_\mu$};
\vertex[dot] (2) at (0,0.7) {};
\vertex (3) at (1.5,0.7) {$A_\mu$};
\vertex (4) at (-1.5,-0.7) {$\psi_{\rm DM}$};
\vertex[dot] (5) at (0,-0.7) {};
\vertex (6) at (1.5,-0.7) {$\psi_{\rm DM}$};
\diagram*{
  (2) -- [photon] (1),
  (3) -- [photon] (2),
  (2) -- [gluon,edge label=\(~h_{\mu\nu}\)] (5),
  (4) -- [fermion] (5),
  (5) -- [fermion] (6)
};
\end{feynman}
\end{tikzpicture}
\label{scalarschannelgraviton}
\end{align}
The $A A h$ vertex is
\begin{align}
\begin{tikzpicture}[baseline=(current bounding box.center)]
\begin{feynman}
\vertex (1) at (-1.0,0.7) {$A_\alpha^a$};
\vertex[dot] (2) at (0,0.7) {};
\vertex (3) at (1.0,0.7) {$A_\beta^b$};
\vertex (4) at (0,-0.7) {};
\diagram*{
  (1) -- [photon] (2),
  (2) -- [photon] (3),
  (2) -- [gluon] (4)
};
\end{feynman}
\end{tikzpicture}
=-\frac{i}{2}\kappa\delta^{ab}\left[(p_1\cdot p_2)C_{\mu\nu\alpha\beta}+D_{\mu\nu\alpha\beta}\right],
\end{align}
where
\begin{align}
C_{\mu\nu,\alpha\beta}
&=
\eta_{\mu\alpha}\eta_{\nu\beta}
+
\eta_{\mu\beta}\eta_{\nu\alpha}
-
\eta_{\mu\nu}\eta_{\alpha\beta},
\\[1ex]
D_{\mu\nu,\alpha\beta}
&=
\eta_{\mu\nu}\,p_{1\beta}p_{2\alpha}
-
\eta_{\mu\beta}\,p_{1\nu}p_{2\alpha}
-
\eta_{\mu\alpha}\,p_{1\beta}p_{2\nu}
-
\eta_{\nu\beta}\,p_{1\mu}p_{2\alpha}
-
\nonumber\\
&\quad
\eta_{\nu\alpha}\,p_{1\beta}p_{2\mu}
+
\eta_{\alpha\beta}
\left(
p_{1\mu}p_{2\nu}
+
p_{1\nu}p_{2\mu}
\right).
\end{align}
The resulting helicity summed matrix element is
\begin{equation}
    \vert \mathcal M_{A} \vert^2=\frac{\kappa^4}{8}\frac{[-t(t+s)][s^2+2t(t+s)]}{s^2}
\end{equation}
where we summed over the 4 possible vector polarization combinations. In the center of mass frame we get
\begin{equation}
    \vert \mathcal M_{A} \vert^2=\frac{\kappa^4s^2}{64}(1-\cos^4\theta).
\end{equation}
There are a total of eight SU(3) gauge bosons, three SU(2)$_L$ ones and one U(1)$_Y$ one to be considered in the final abundance calculation. 

For following calculations it is convenient to define the stripped matrix elements ${|\mathcal{\widetilde M}|^2}$ as follows
\begin{equation}
    \label{eq:strippedMEdef}
    |\mathcal{M}|_i^2=\frac{\kappa^4s^2}{256}w^\prime_i{|\widetilde{\mathcal{M}}|_i^2},
\end{equation}
where the weights $w^\prime_i$ are calculated from the matrix element prefactors. When computing the dark matter abundance using the Boltzmann equation, a prefactor of 1/2 to account for the two identical incoming particles in the case of vector and scalar particles appears. Also, a final factor of 2 for production of dark matter particle antiparticle pair, not just one particle, appears. We therefore define the Boltzmann equation corrected weights $w_i$ that also include these factors. The Boltzmann equation corrected weights $w_i$ are given in Table~\ref{tab:angular_kernels}.
\begin{table}[h!]
    \centering
    \renewcommand{\arraystretch}{1.3}
    \begin{tabular}{lcc}
        \hline\hline
        Interaction $i$
        & ${|\mathcal{\widetilde M}|^2}$
        & $w_i$ \\
        \hline
        fermion + fermion (EC)
        & $1+\cos^2\theta$ & $9$ \\

        fermion + fermion (s-channel graviton exchange)
        & $1-3\cos^2\theta+4\cos^4\theta$ & $1$ \\

        fermion + fermion (EC-graviton interference)
        & $3\cos^2\theta-1$ & $6$ \\

        scalar + scalar (s-channel graviton exchange)
        & $\cos^2\theta-\cos^4\theta$ & $2$ \\

        vector + vector (s-channel graviton exchange)
        & $1-\cos^4\theta$ & $4$ \\
        \hline\hline
    \end{tabular}
    \caption{Stripped matrix elements and their weights for all interactions.}
    \label{tab:angular_kernels}
\end{table}

\subsection{Phase space distribution}
\label{sec:phase}

While the total fermion-initiated abundances produced by the EC interaction and the s-channel graviton exchange simply add up, the interference has an effect on the phase space distribution of the produced dark matter. In this section we compute the Boltzmann equation collision term
\begin{equation}
    C(q)=\frac{1}{2E}\int\frac{d^3 p_1}{(2\pi)^32E_1}\int\frac{d^3 p_2}{(2\pi)^32E_2}\int\frac{d^3 l}{(2\pi)^32E_l}(2\pi)^4\delta^{(4)}(p_1+p_2-q-l)|\mathcal M|^2f(p_1)f(p_2),
    \label{eq:intdef}
\end{equation}
where $q$ is the four momentum of the observed outgoing dark matter particle, $l$ the four momentum of the other produced dark matter particle and $p_1,p_2$ the four momenta of the incoming SM particles. Following \cite{Shaposhnikov:2020aen}, we will assume that the SM particles form a thermalized bath at the time of production and so their distributions are described by the Fermi-Dirac distribution
\begin{equation}
    f_{1,2}=\frac{1}{e^{E_{1,2}/T}+1}.
\end{equation}
The key point is that for a given $q$, the center-of-mass system is boosted relative to the thermal bath and thus the angular integration gains additional weight terms which make the contribution of the interference term nonzero. The detailed calculation is presented in Appendix \ref{sec:integral}. 

The scalar and vector production channels remain unaffected by the EC interaction, while the fermion sourced dark matter energy distribution slightly changes. Let us define the dimensionless variable $y=E/T$ and a weighted function
\begin{equation}
    J_w(y)=\xi\left(9C_{\rm EC}(y)+C_{\rm G}(y)+6 C_{\rm int}(y)\right),~~~\xi=\frac{16}{\kappa^4 T^5},
    \label{eq:Jdef}
\end{equation}
where $C_{\rm EC}$, $C_{G}$ and $C_{\rm int}$ are computed using the stripped matrix elements.

\begin{figure}[H]
\begin{center}
\includegraphics[width=.49\linewidth]{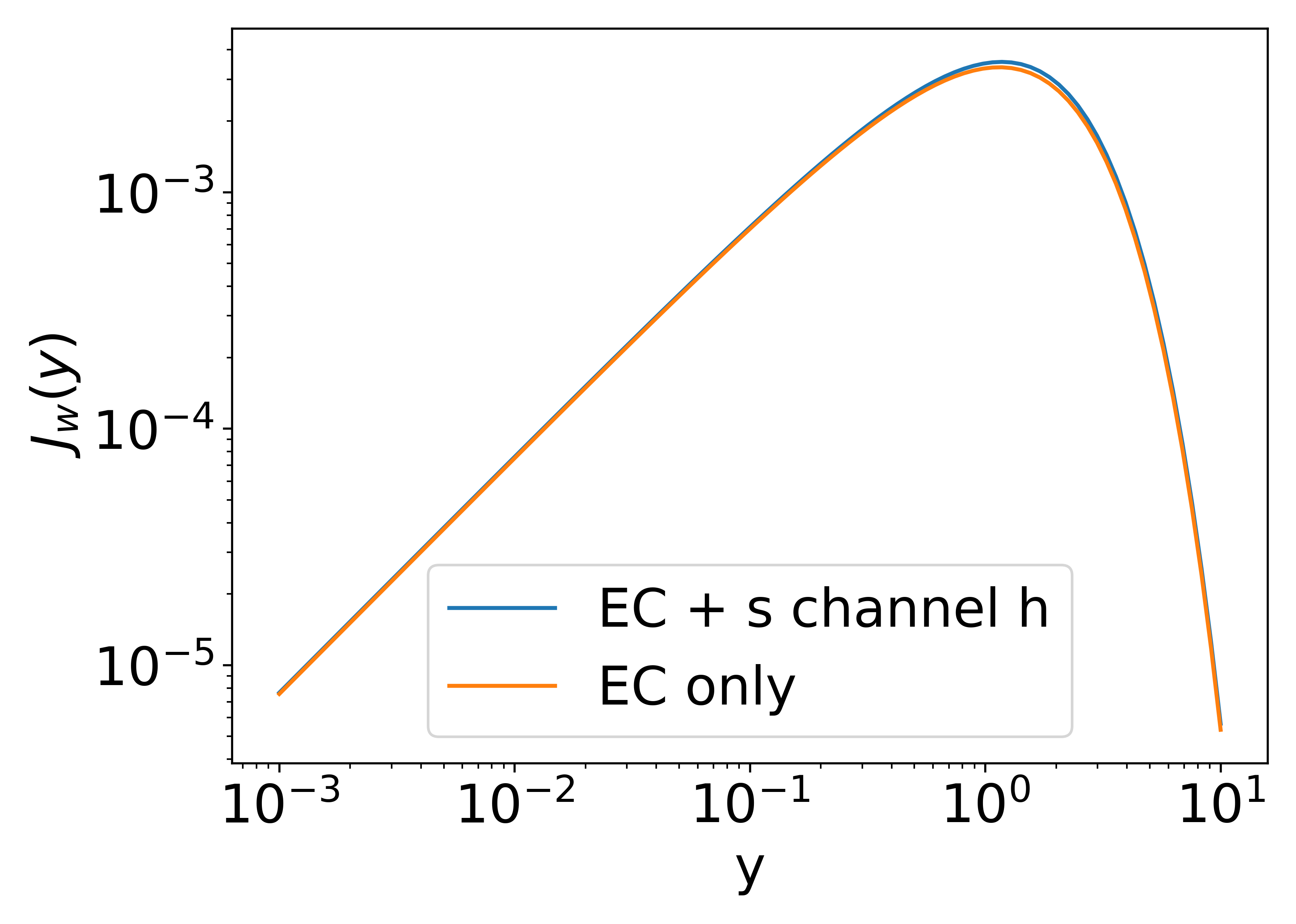}
\includegraphics[width=.49\linewidth]{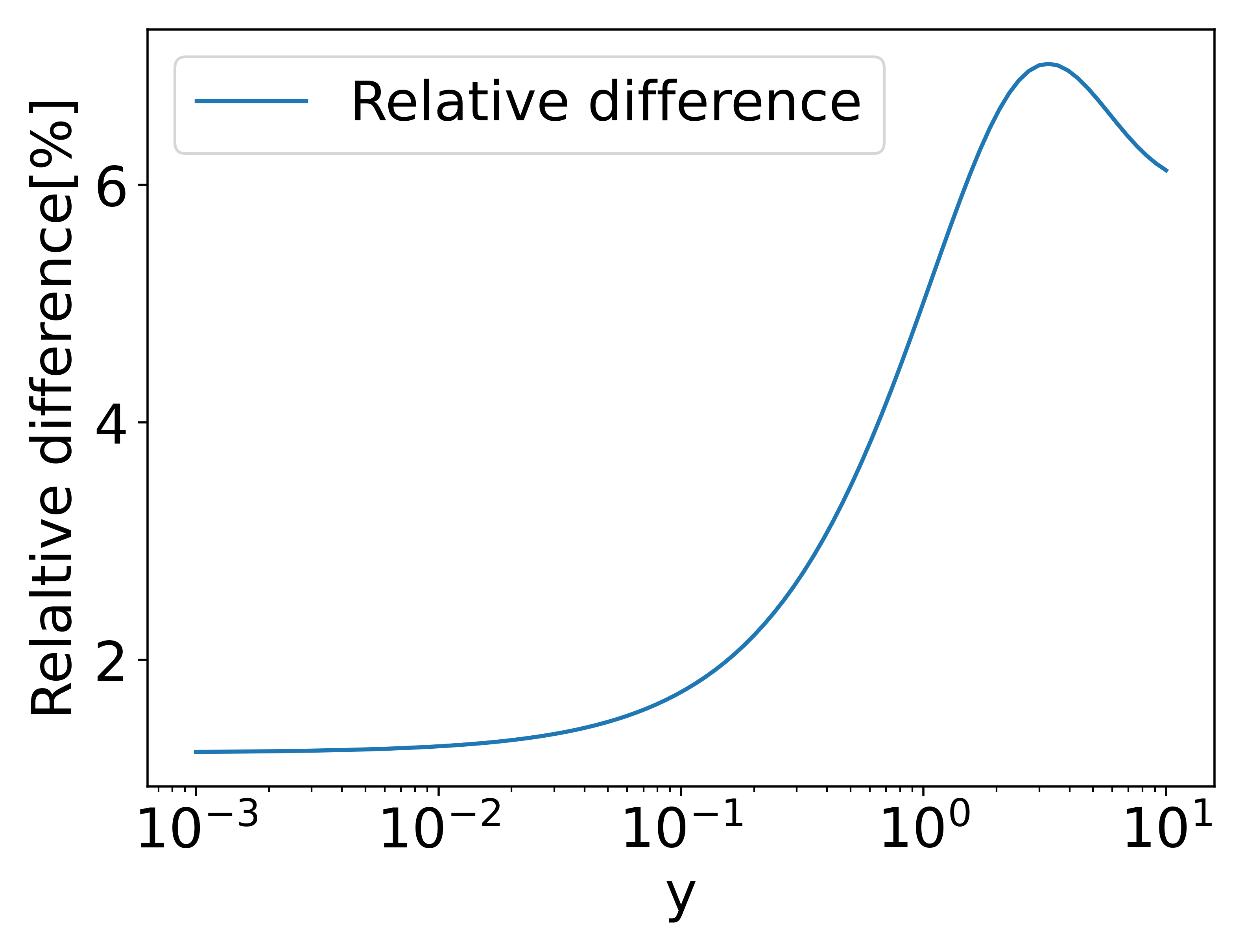}
\caption{
The dimensionless weighted collision term $J_w(y)$ for pure EC interaction and with the effect of the s-channel graviton exchange (left). The relative difference between the two curves (right).}
\label{figPureEC}
\end{center}
\end{figure}

The results of numerical integration are presented in Fig.~\ref{figPureEC}. Compared to the EC interaction only case, the effect of the s-channel graviton exchange is more pronounced towards the high energy end of the spectrum, resulting in about 7\% difference at $y=3.27$. The total fermion-initiated instantaneous production rate integral
\begin{equation}
   \int_0^\infty y^2J_w(y)dy
\end{equation}
is not affected by the interference term
\begin{equation}
    \label{eq:int_total}
    \int_0^\infty y^2C_{\rm int}(y)dy=0
\end{equation}
and therefore the total fermion-initiated dark matter abundance at the end is just a sum of those from the two processes.

\subsection{Dark matter abundance}
\label{sec:abundance}

Now let us evaluate the dark matter abundance. 
As we have shown, there are three production channels. Production from fermion channels have been mentioned in the previous subsection.
As for the scalar and gauge boson channels, we have to consider Bose-Einstein distribution
\begin{equation}
    f_{1,2}=\frac{1}{e^{E_{1,2}/T}-1}
\end{equation}
instead of Fermi-Dirac one (see the comment at the end of Appendix \ref{sec:integral}). The calculation is otherwise identical to the previous subsection. The integral
\begin{equation}
    R=\frac{T^3}{2\pi^2}\int_0^\infty y^2C(y)dy
\end{equation}
can be evaluated analytically, see Appendix \ref{sec:integral} for the calculation. The result per real scalar/vector field or per chiral fermion can be summarized as
\begin{equation}
    R_i=\frac{T^8\pi^3}{86400 M_{\rm pl}^4}F_iw_i\langle{|\mathcal{\widetilde M}|^2}\rangle_\phi,
\end{equation}
where
\begin{equation}
    \langle{|\mathcal{\widetilde M}|^2}\rangle_\phi=\frac{1}{2}\int_{-1}^1d\cos\theta{|\mathcal{\widetilde M}|^2},
\end{equation}
$F_i=1$ for bosons and $F_i=49/64$ for fermions. Concretely we get
\begin{equation}
    R_{\rm scalar}=\frac{T^8\pi^3}{M_{\rm pl}^4}\frac{1}{324000},~~~ R_{\rm vector}=\frac{T^8\pi^3}{M_{\rm pl}^4}\frac{1}{27000},~~~~ R_{\rm fermion}=\frac{T^8\pi^3}{M_{\rm pl}^4}\frac{49}{432000}.
\end{equation}
Taking the respective number of species and summing together, we get
\begin{equation}
    R_{\rm total}=\frac{T^8\pi^3}{M_{\rm pl}^4}\frac{7207}{1296000}\approx 0.1724\frac{T^8}{M_{\rm pl}^4}. 
\end{equation}
For reference, if there were no EC interaction, the fermion contribution $R_{\rm fermion}$ should be divided by a factor $16$ (see Eq.~(\ref{EC_ratio})) and the result would be\footnote{Ref.~\cite{Clery:2021bwz} presents a different result. The discrepancy is caused by us counting the SM fermions as 45 chiral species, whereas Ref.~\cite{Clery:2021bwz} treats them as Dirac fermions.}
\begin{equation}
     R^{\rm (noEC)}_{\rm total}=\frac{T^8\pi^3}{M_{\rm pl}^4}\frac{16087}{20736000}\approx 0.0241\frac{T^8}{M_{\rm pl}^4}.
\end{equation}

The dark matter number density $n$ evolves according to 
\begin{equation}
    \dot{n}+3Hn=R_{\rm total}(T),
\end{equation}
where $H = \dot a/a$ denotes the Hubble parameter with $a(t)$ being the cosmic scale factor.
Multiplying both sides by $a^3$ we can write the solution
\begin{equation}
    n(t)a(t)^3=\int_{t_i}^t a^3 R_{\rm total}(T)dt.
\end{equation}
In a standard reheating scenario, in which the inflaton oscillation behaves as non-relativistic matter and the reheating is completed at the temperature $T_{\rm RH}$, it is found that the dark matter production is dominated at $T\simeq T_{\rm RH}$.\footnote{Here we assumed that dark matter mass $m_{\rm DM}$ is smaller than $T_{\rm RH}$. Otherwise, the production is dominated at $T\sim m_{\rm DM}$. In such a case, the following estimate for the dark matter abundance should be multiplied by a factor $\sim (T_{\rm RH}/m_{\rm DM})^4$ for $m_{\rm DM} > T_{\rm RH}$~\cite{Ema:2019yrd}.}
The number density around $T\simeq T_{\rm RH}$ is estimated as
\begin{equation}
    n(T_{\rm RH})\approx\frac{R_{\rm total}(T_{\rm RH})}{3H(T_{\rm RH})}.
\end{equation}
The number-to-entropy density ratio, $Y_\infty = n/s$ with $s=(2\pi^2 g_{*s}/45)T^3$, is fixed at this epoch. It is given by
\begin{equation}
    Y_\infty\approx2.485\times 10^{-20}\left(\frac{T_{\rm RH}}{10^{13} \rm GeV}\right)^3.
\end{equation}
Without the EC interaction we get
\begin{equation}
    Y^{\rm (no EC)}_\infty\approx3.466\times 10^{-21}\left(\frac{T_{\rm RH}}{10^{13} \rm GeV}\right)^3.
\end{equation}
In terms of the dark matter density parameter, they are written as
\begin{equation}
    \Omega_c h^2 \approx 0.068\left(\frac{m_{\rm DM}}{10^{10}\,{\rm GeV}}\right)\left(\frac{T_{\rm RH}}{10^{13}\,\rm GeV}\right)^3,
\end{equation}
where $m_{\rm DM}$ is the dark matter mass. This expression is applicable when $m_{\rm DM} < T_{\rm RH}$.
If we would neglect the EC interaction, we instead have
\begin{equation}
    \left[\Omega_c h^2\right]^{\rm (noEC)} \approx 9.5\times 10^{-3}\left(\frac{m_{\rm DM}}{10^{10}\,{\rm GeV}}\right)\left(\frac{T_{\rm RH}}{10^{13}\,\rm GeV}\right)^3.
\end{equation}

It should be noticed that we ignored the gravitational production from the inflaton coherent oscillation (or inflaton annihilation)~\cite{Ema:2015dka,Ema:2016hlw,Ema:2018ucl,Ema:2019yrd}, since it significantly depends on the inflation model parameters such as inflaton mass and inflation energy scale.

\section{Effect of the Holst term}
\label{sec:holst}

\subsection{Fermion interactions in the presence of Holst term}

Presence of torsion allows us to consider one more nontrivial scalar linear in curvature. Besides the usual parity even Ricci scalar $R=R_{\mu\nu\alpha\beta}g^{\mu\alpha}g^{\nu\beta}$, we can introduce a parity odd one called the Holst term~\cite{Holst:1995pc}
\begin{equation}
    \widetilde{R}=\frac{1}{2}R_{\mu\nu\alpha\beta}\epsilon^{\mu\nu\alpha\beta}.
\end{equation}
When torsion is zero the Holst term vanishes, but in the presence of torsion it can be written as \cite{Nieh:1981ww} (up to a total derivative)
\begin{equation}
    \widetilde{R}=\epsilon^{\mu\nu\alpha\beta}\tensor{K}{_\mu^\lambda_\alpha}\tensor{K}{_\lambda_\nu_\beta}.
\end{equation}
The gravitational sector of the action becomes
\begin{equation}
    S_{\rm G}=-\int d^4x\sqrt{-g}\frac{2}{\kappa^2}\left(R-\frac{1}{\gamma}\widetilde{R}\right).
\end{equation}
This action has many applications as it is usually considered the starting point for loop quantization of gravity. The dimensionless parameter $\gamma$ is called the Barbero-Immirzi parameter~\cite{BarberoG:1994eia,Immirzi:1996di}.

When added to the pure Dirac field action the Holst term only modifies the EC coupling. However, if we introduce a nonminimal massless fermion Lagrangian of the form
\begin{equation}
    {S_\psi}=\int d^4x\sqrt{-g}\frac{i}{2}\left[(1-i\alpha)\bar{\psi}\gamma^\mu D_\mu\psi-(1+i\alpha)D_\mu\bar{\psi}\gamma^\mu\psi\right],
    \label{eq:nonminFermion}
\end{equation}
where $\alpha$ is real, we get a set of three interactions~\cite{Freidel:2005sn,Perez:2005pm}
\begin{align}
        \label{eq:Holstinteraction}
        {S}_{4\psi}=&\int d^4x\sqrt{-g}\left[\frac{3\kappa^2}{64}\frac{\gamma^2}{\gamma^2+1}\left(A^\mu A_\mu+\frac{2\alpha}{\gamma}A^\mu V_\mu-\alpha^2V^\mu V_\mu\right)\right],\\
        V_\mu=&\sum_f V_{f\mu},~~A_\mu=\sum_f A_{f\mu},
\end{align}
where $V^\mu=\bar{\psi}\gamma^\mu\psi$ is the vector current and $A^\mu=\bar{\psi}\gamma_5\gamma^\mu\psi$ is the axial vector current. We assume the constant $\alpha$ to be the same for all fermion species. 
There are three types of interactions: $AA$, $AV$ and $VV$ terms. 
The first term is just the usual EC interaction with the coupling suppressed by $\gamma^2/(1+\gamma^2)$. In the limit $\gamma\rightarrow\infty, \alpha\rightarrow0$ we recover the original EC interaction, while taking $\gamma\rightarrow0$ forces vanishing torsion and the interactions disappear. 

For chiral SM fermions we can simplify the calculation by writing
\begin{equation}
    A_{f\mu}^{\rm SM}=\sigma_fV_{f\mu}^{\rm SM},
\end{equation}
where $\sigma_f=-1$ for right-handed fermion and $\sigma_f=1$ for left-handed one. Then the interaction for a single chiral fermion species \eqref{eq:Holstinteraction} becomes
\begin{equation}
    {S}_{4\psi}=\int d^4x\sqrt{-g}\left[\frac{3\kappa^2}{32}\frac{\gamma^2}{\gamma^2+1}\left(V_f^\mu(\sigma_f+r) A_\mu+ (r\sigma_f-\alpha^2)V^\mu_f V_\mu\right)\right],~~~r=\frac{\alpha}{\gamma}.    
\end{equation}
So we get a helicity dependent $VV$ and $AV$ interaction. For the $VV$ interaction, we get an interference term with the s-channel graviton exchange of the form\footnote{Since there are no $\gamma_5$ in the $VV$ interaction, we can treat the SM fermions as Dirac and divide the resulting squared amplitudes by 2.}
\begin{equation}
    {\rm Tr}\left[\slashed{p_2}\gamma_\mu\slashed{p_1}\gamma_\rho (p_1-p_2)_\sigma\right]D^{\rho\sigma\alpha\beta}{\rm Tr}\left[\slashed{p_3}\gamma^\mu\slashed{p_4}\gamma_\alpha (p_3-p_4)_\beta\right].
\end{equation}
Now there is no $\gamma_5$ so the result will be nonzero unless some precise cancellation occurs. Evaluating the traces gives us
\begin{equation}
    2\Re(\mathcal M_{VV}\mathcal M_{\rm G}^*)\propto s^2\cos^3\theta.
\end{equation}
This is an odd function of $\cos\theta$ so after the full angle integration we get zero. The squared amplitude per one chiral SM fermion summed over the dark matter helicities is
\begin{equation}
    |\mathcal{M}_{VV}^\sigma|^2=\frac{9\kappa^4 \Lambda^2}{256}(\sigma r-\alpha^2)^2(t^2+u^2),~~~\Lambda=\frac{\gamma^2}{\gamma^2+1}.
\end{equation}
In terms of the scattering angle this becomes
\begin{equation}
    |\mathcal{M}_{VV}^\sigma|^2=\frac{9\kappa^4 s^2 \Lambda^2}{512}(\sigma r-\alpha^2)^2(1+\cos^2\theta).
\end{equation}
The trace term corresponding to the axial dark matter current is
\begin{equation}
    {\rm Tr}\left[\slashed{p_3}\gamma_5\gamma^\mu\slashed{p_4}\gamma_5\gamma^\nu\right]={\rm Tr}\left[\slashed{p_3}\gamma^\mu\slashed{p_4}\gamma^\nu\right].
\end{equation}
This is exactly the same as for the vector current, so we can immediately write down the resulting amplitude
\begin{equation}
    |\mathcal{M}_{AV}^\sigma|^2=\frac{9\kappa^4 \Lambda^2}{256}(\sigma + r)^2(t^2+u^2).
\end{equation}
The $VV$-$AV$ interference term is
\begin{equation}
    2\Re(\mathcal{M}_{AV}^\sigma\mathcal{M}_{VV}^{\sigma*})=\frac{9\kappa^4\Lambda^2}{128}\sigma(\sigma r-\alpha^2)(\sigma+r)(u^2-t^2).
\end{equation}
This is linear function of $\cos\theta$
\begin{equation}
    2\Re(\mathcal{M}_{AV}^\sigma\mathcal{M}_{VV}^{\sigma*})=\frac{9\kappa^4s^2\Lambda^2}{128}\sigma(\sigma r-\alpha^2)(\sigma+r)\cos\theta.
\end{equation}
and so it vanishes after the full angle integration. 

Finally we have to compute the interference between the $AV$ interaction and the s-channel graviton exchange. For the case of Dirac SM fermions, we get an interference term with the s-channel graviton exchange of the form
\begin{equation}
    {\rm Tr}\left[\slashed{p_2}\gamma_\mu\slashed{p_1}\gamma_\rho (p_1-p_2)_\sigma\right]D^{\rho\sigma\alpha\beta}{\rm Tr}\left[\slashed{p_3}\gamma_5\gamma^\mu\slashed{p_4}\gamma_\alpha (p_3-p_4)_\beta\right],
\end{equation}
The first trace has no $\gamma_5$ that would lead to a Levi-Civita tensor, so ultimately it is just a sum of terms involving metric contractions. Contracting with the graviton propagator is again just a set of metric contractions, so finally we must end up with just one nonzero term proportional to
\begin{equation}
    \mathcal M_{AV}\mathcal M_{\rm G}^*\propto\epsilon_{\alpha\beta\mu\nu}p_1^\alpha p_2^\beta p_3^\mu p_4^\nu.
\end{equation}
Because the momentum conservation forces $p_1+p_2=p_3+p_4$, the result is zero. However, this does not apply to chiral SM fermions because of the addition $\gamma_5$ introduced by the projectors. Taking chiral fermions instead, we get
\begin{equation}
    2\Re(\mathcal M_{AV}\mathcal M_{\rm G}^*)\propto(1+\sigma r)(1-3\cos^2\theta).
\end{equation}
This again vanishes after the full angular integration.

\subsection{Dark matter abundance}

Let us first consider the case $\alpha=0$. This corresponds to simply decreasing the EC interaction coupling. Keeping the minimal fermion, the relation \eqref{eq:Jdef} becomes
\begin{equation}
      J_w(y)=\xi\left(9\Lambda^2C_{\rm EC}(y)+C_{\rm G}(y)+6\Lambda C_{\rm int}(y)\right).
\end{equation}
Taking lower values of $\Lambda$ brings the strength of the EC channel closer to that of the s-channel graviton exchange, resulting in additional deformation of the distribution. At $\Lambda=1/3$ their respective prefactors become equal. As the production is dominated by the EC interaction lowering $\Lambda$ reduces it substantially. In the case $\Lambda=0$ we are left with only the s-channel graviton exchange.
\begin{figure}[H]
\begin{center}
\includegraphics[width=0.49\linewidth]{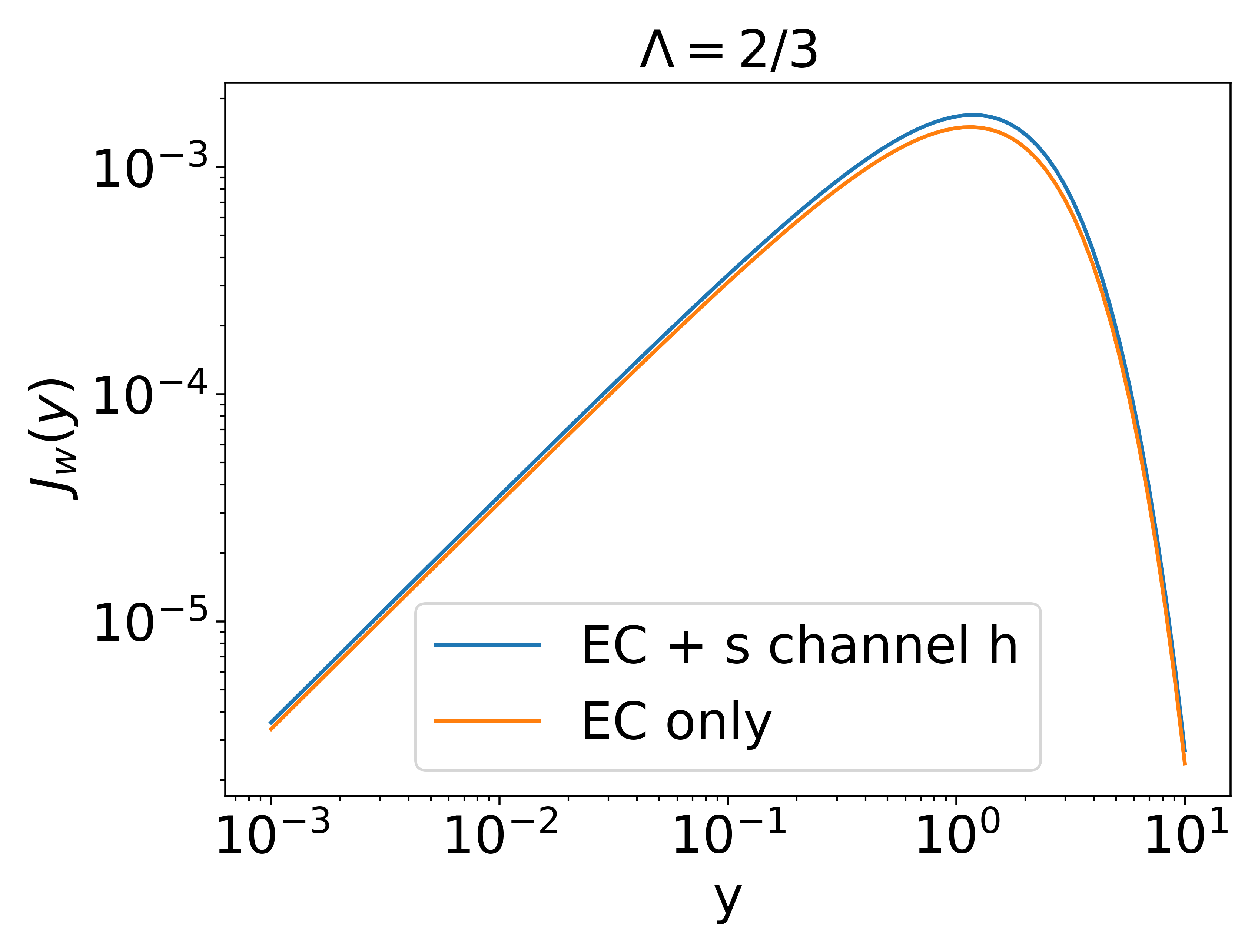}
\includegraphics[width=0.49\linewidth]{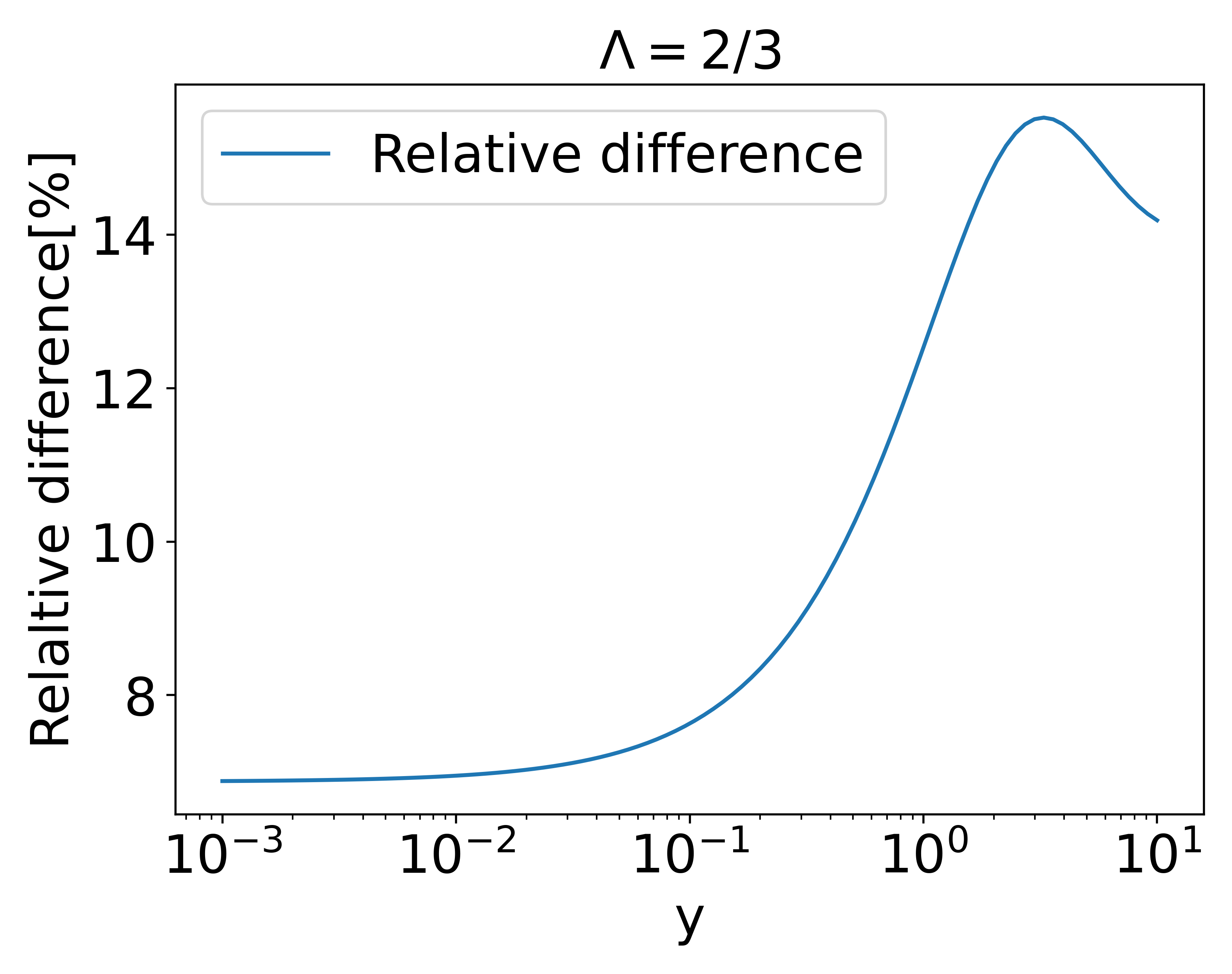}
\includegraphics[width=0.49\linewidth]{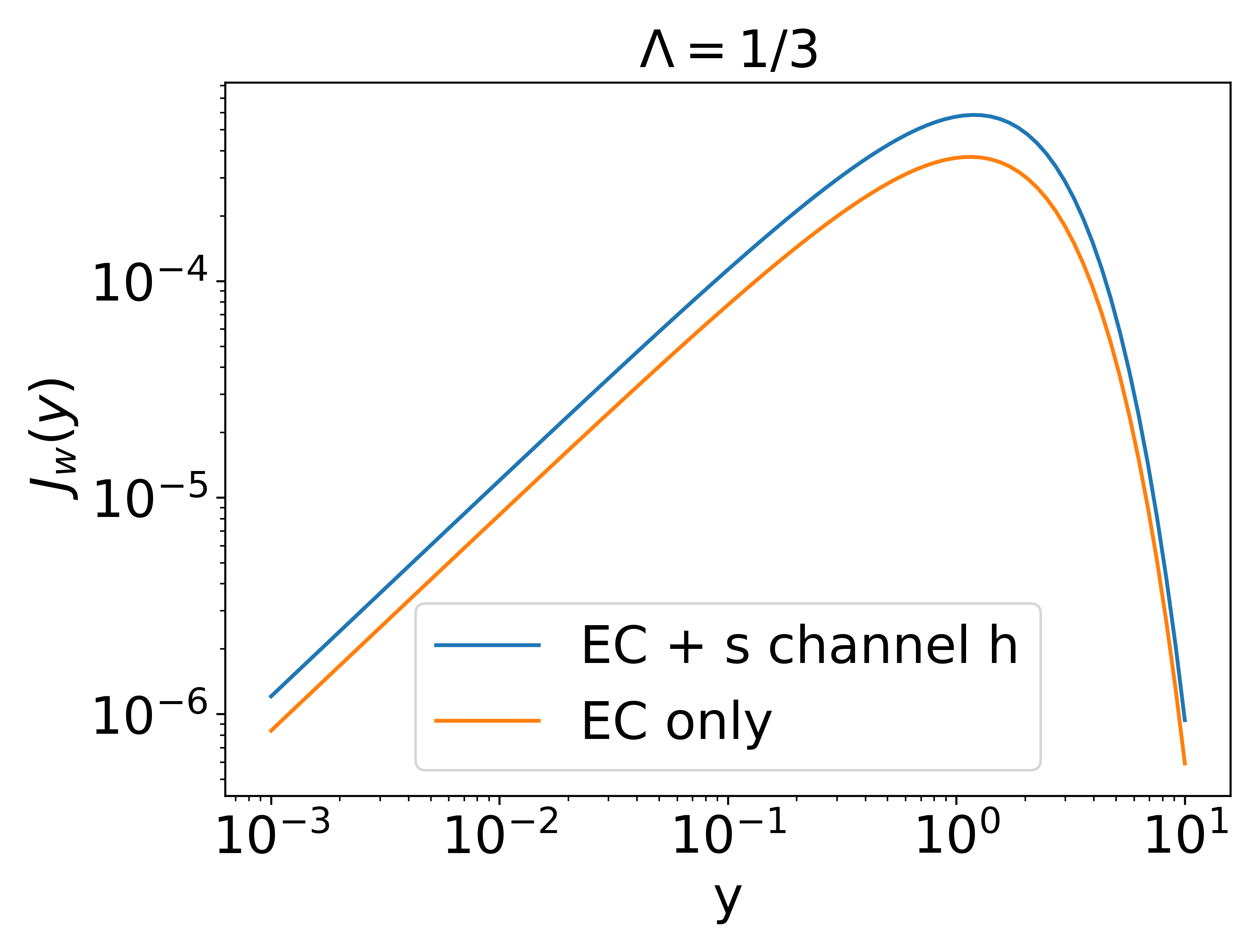}
\includegraphics[width=0.49\linewidth]{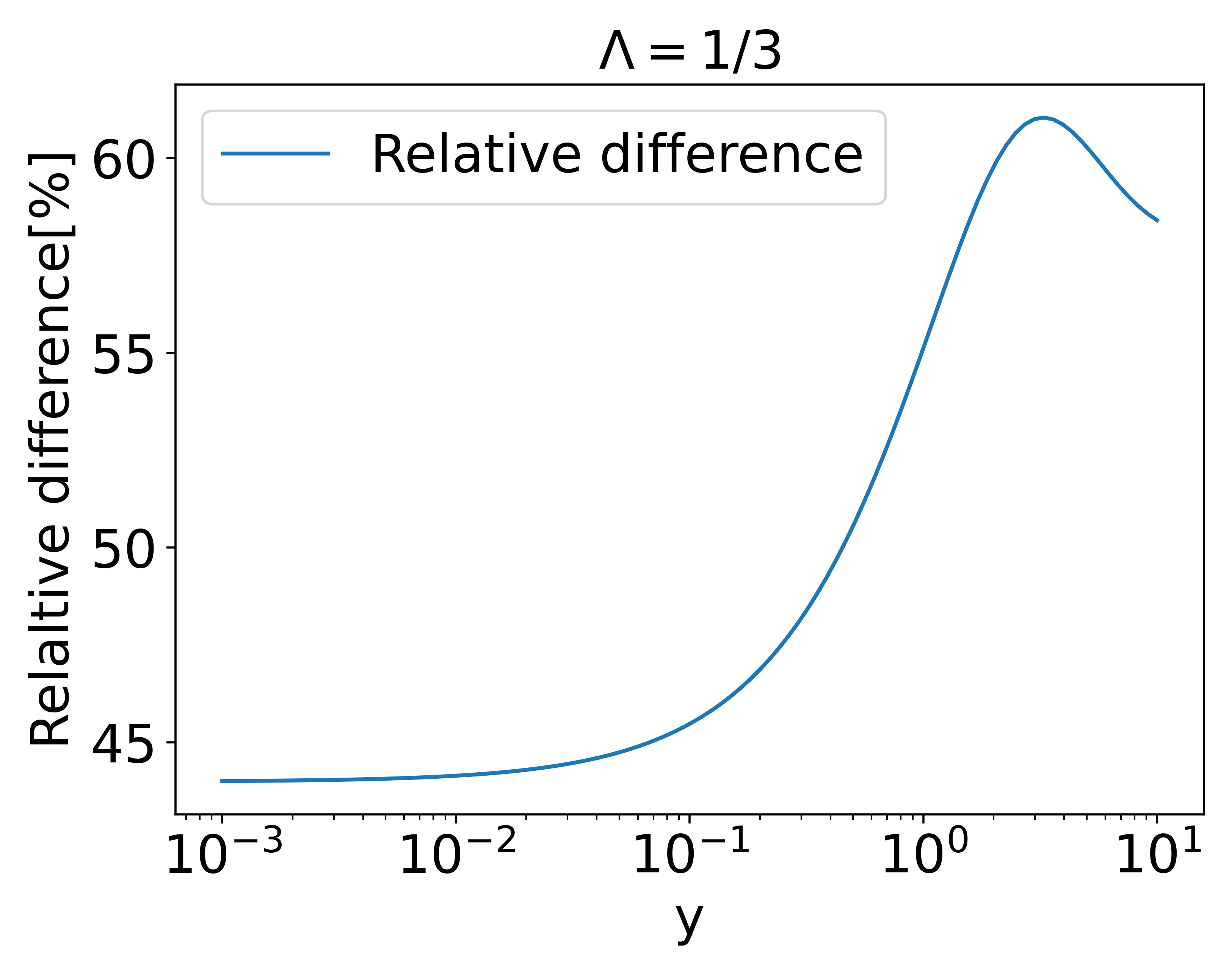}
\caption{The dimensionless weighted collision term $J_w(y)$ for Holst term modified EC interaction considering the minimal $\alpha=0$ fermion. Upper two panels show the absolute magnitude and relative difference to EC channel only for $\Lambda=2/3$. Two bottom panels show the same for $\Lambda=1/3$. We do not take the scalar and vector species into account in this plot.}
\label{figHolst}
\end{center}
\end{figure}

Continuing with the full $\alpha\neq0$ interaction, the stripped matrix elements and weights for \eqref{eq:strippedMEdef} are given by Table 2. 
\begin{table}[th!]
    \centering
    \begin{tabular}{lcc}
        \hline
        Interaction $i$ &
        $\widetilde{|\mathcal{M}|}^{2}$ &
        $w_i$ \\
        \hline

        fermion (s-channel graviton)
        & $1-3\cos^2\theta+4\cos^4\theta$
        & $1$ \\

        fermion (Holst $AV$)
        & $1+\cos^2\theta$
        & $9\Lambda^2(\sigma_f+r)^2$ \\

        fermion (Holst $VV$)
        & $1+\cos^2\theta$
        & $9\Lambda^2(\sigma_f r-\alpha^2)^2$ \\

        fermion ($AV$--$VV$ interference)
        & $\cos\theta$
        & $36\Lambda^2\sigma_f(\sigma_f+r)
           (\sigma_f r-\alpha^2)$ \\

        fermion (Graviton--$AV$ interference)
        & $3\cos^2\theta-1$
        & $6\Lambda(1+\sigma_f r)$ \\

        fermion (Graviton--$VV$ interference)
        & $-\cos^3\theta$
        & $12\Lambda(\alpha^2-\sigma_f r)$ \\

        scalar (s-channel graviton)
        & $\cos^2\theta-\cos^4\theta$
        & $2$ \\

        vector (s-channel graviton)
        & $1-\cos^4\theta$
        & $4$ \\

        \hline
    \end{tabular}
    \caption{Stripped matrix elements and their weights
    for all interactions including the Holst term. The first column shows classification of initial particle and its interactions through which it pair-annihilates into dark matter.}
    \label{tab:holst_matrix_elements}
\end{table}
The overall abundance becomes
\begin{equation}
      R^{\rm (Holst)}_{\rm total}(\gamma,\alpha)=\frac{T^8\pi^3}{M_{\rm pl}^4}\left[\frac{16087}{20736000}+\frac{99225}{20736000}\left(\frac{\gamma^2}{1+\gamma^2}\right)^2\left(1+2r^2+\alpha^4+\frac{2}{15}\frac{\alpha}{\gamma}(1-\alpha^2)\right)\right].
\end{equation}
where we considered 21 right handed fermions and 24 left handed ones. Note that the additional contribution compared to pure GR cannot be negative for any choice of $\gamma,\alpha$. 
In terms of the dark matter density parameter, we obtain
\begin{align}
    \left[\Omega_c h^2\right]^{\rm (Holst)} \approx & 9.5\times 10^{-3}\left(\frac{m_{\rm DM}}{10^{10}\,{\rm GeV}}\right)\left(\frac{T_{\rm RH}}{10^{13}\,\rm GeV}\right)^3 \nonumber \\
    &\times\left[1+\frac{99225}{16087} \left(\frac{\gamma^2}{1+\gamma^2}\right)^2\left(1+2r^2+\alpha^4+\frac{2}{15}\frac{\alpha}{\gamma}(1-\alpha^2)\right) \right].
\end{align}

\section{Conclusions}
\label{sec:conc}

We have improved the calculation of dark matter production in EC gravity by including the s-channel graviton exchange. The minimal EC theory introduces one universal axial-axial interaction between all possible fermion pairs. The s-channel diagram produces an interference term that modifies the angular distribution of the products but vanishes when integrated over the full angle. Taking into account all Standard model species including scalars and vector bosons, the overall abundance is increased by a factor of approximately $7.17$ compared to the pure GR case without torsion. Although the interference does not contribute to the total fermion-initiated dark matter abundance, it modifies the phase space distribution. We computed the Boltzmann equation collision term for fermion initiated dark matter production and found that most strongly amplified part is the high energy tail, which is boosted by about 7\% at the maximally enhanced energy.

It is possible to modify the coupling by including the Holst term in the action. The Holst term introduces two additional interactions, an axial-vector one and a vector-vector one. Neither the EC interaction nor the axial-vector and vector-vector interactions from the Holst term interfere with the s-channel graviton exchange in way that would affect the total cross section. The resulting abundances add up. These three types are all the interactions arising from integrating out torsion from leading local fermion-torsion couplings~\cite{Diakonov:2011fs}. In the case of minimal fermion action the presence of the Holst term can only lower the total cross section. If we include a non-minimal fermion action \eqref{eq:nonminFermion}, the production can be increased compared to the pure EC theory. 

\section*{Acknowledgment}

This work was supported by World Premier International Research Center Initiative (WPI), MEXT, Japan.
This work was also supported by JSPS KAKENHI (Grant Number 24K07010 [KN], 26K00695 [KN], 26H00403 [KN]).

\appendix

\section{Matrix elements for general massive fermions}  \label{sec:massive}
In this Appendix we provide a full formula for the squared matrix element keeping the dark matter mass ($m_{\rm DM}$) finite. We define the usual variable
\begin{equation}
    \beta=\sqrt{1-\frac{4 m_{\rm DM}^2}{s}}.
\end{equation}
For the Holst term VV interaction we get
\begin{equation}
    |\mathcal{M}_{VV}^\sigma|^2=\frac{9\kappa^4 s^2 \Lambda^2}{512}(\sigma r-\alpha^2)^2\left[2+\beta^2(\cos^2\theta-1)\right].
\end{equation}
For the Holst term AV interaction we get
\begin{equation}
    |\mathcal{M}_{AV}^\sigma|^2=\frac{9\kappa^4 s^2 \Lambda^2}{512}(\sigma + r)^2\beta^2\left(1+\cos^2\theta\right).
\end{equation}
The interference between them becomes
\begin{equation}
    2\Re(\mathcal{M}_{AV}^\sigma\mathcal{M}_{VV}^{\sigma*})=\frac{9\kappa^4s^2\Lambda^2}{128}\sigma(\sigma r-\alpha^2)(\sigma+r)\beta\cos\theta.
\end{equation}
This still vanishes after the full angular integration. For the s channel graviton exchange we get
\begin{equation}
     |\mathcal{M}_{G}^\sigma|^2=\frac{\kappa^4 s^2 \beta^2}{512}\left[1+\cos^2\theta-4\beta^2\cos^2\theta(1-\cos^2\theta)\right].
\end{equation}
The interferences become
\begin{equation}
    2\Re(\mathcal M_{VV}\mathcal M_{\rm G}^*)=-\frac{3\kappa^4\Lambda s^2}{128}(\alpha^2-\sigma r)\,\beta\cos\theta\left[1-\beta^2+\beta^2\cos^2\theta\right],
\end{equation}
\begin{equation}
    2\Re(\mathcal M_{AV}\mathcal M_{\rm G}^*)=-\frac{3\kappa^4\Lambda s^2}{256}(1+\sigma r)\,\beta^2(1-3\cos^2\theta).
\end{equation}
These again fully vanish after the angular integration.
\section{Angular dependence of dark matter production}
\label{sec:ang}

We have shown that it is not possible to cancel out the s-channel graviton exchange dark matter production using the EC interaction, but it is at least possible to modify the angular distribution of the created dark matter particles. We can use the Legendre polynomial decomposition of the squared matrix elements
\begin{equation}
    \vert \mathcal M\vert^2=\sum_{n=0}^{\infty}a_n P_n(\cos\theta),
\end{equation}
where $P_n$ are the Legendre polynomials. Since these form an orthogonal basis, i.e.
\begin{equation}
    \int_{-1}^{1}P_n(x)P_m(x)dx=0~~~\text{when}~~~n\neq m
\end{equation}
and $P_0(x)=1$, we immediately see that the total reaction rate is only determined by $a_0$. The higher terms in the expansion only modify the angular distribution of the products, but not the total rate. 

From this point of view, the cancellation of the interference terms between the EC theory and scalar or vector interactions is different from that with gravity. The scalar interactions result in $a_n=0~\forall n$, while the vector interactions give $a_0=0,~a_1\neq0,~a_n=0~(n\geq2)$. So in principle it is possible to remove the quadrupole moment of the produced dark matter angular distribution by tuning the EC coupling constant. 

For the combined s-channel graviton and EC squared matrix element we get the nonzero coefficients
\begin{align}
     a_0&=\frac{\kappa^{4} \left(45 \Lambda^2 s^{2} + 3 s^{2}\right)}{1920},\\
     a_2&=\frac{\kappa^{4} \left(63 \Lambda^2 s^{2} + 126 \Lambda s^{2} + 3 s^{2}\right)}{5376},\\
     a_4&=\frac{\kappa^{4} s^{2}}{560},
\end{align} 
where $\Lambda$ is a coefficient that multiplies the standard EC interaction constant. $\Lambda=1$ corresponds to the usual EC theory. The choice 
\begin{equation}
    \Lambda_-=-1-\frac{2\sqrt{105}}{21},~~~\Lambda_+=-1+\frac{2\sqrt{105}}{21},
\end{equation}
sets $a_2=0$. As both values are negative, cancellation of the quadrupole moment is not achievable by introducing the Holst term with a real coupling $\gamma$ as was done in this work.

\section{Phase space integral}
\label{sec:integral}

In this section we present the detailed calculation of the phase space integral \eqref{eq:intdef}. Here we consider all particles to be massless. Reverse reactions and Pauli blocking are neglected. The Lorentz invariant measure is
\begin{equation}
    d\Pi_i=\frac{d^3 p_i}{(2\pi)^32E_i}.
\end{equation}
Let us evaluate it in the center-of-mass frame. First, we parameterize the $p_1,p_2$ sector. We denote the total momentum of the system by $P$. Integrating over the spatial part of the delta function we get
\begin{equation}
    d\Phi(p_1,p_2,P)=d\Pi_1d\Pi_2(2\pi)^4\delta(P^0-E_1-E_2)\delta^{(3)}(P-p_1-p_2)=\frac{1}{16\pi^2}\frac{d^3 p_1}{E_1 E_2}\delta(P^0-E_1-E_2)
\end{equation}
and in the center-of-mass frame $p_1=-p_2$. Going to spherical coordinates we get
\begin{equation}
   d\Phi(p_1,p_2,P_*)=\frac{d\Omega}{16\pi^2}\int dp_1\frac{p_1^2}{E_1 E_2}\delta(\sqrt{s}-E_1-E_2).
\end{equation}
Using 
\begin{equation}
    \delta(f(p))=\sum_{p_*}\frac{\delta(p-p_*)}{\vert f^{\prime}(p_*)\vert},~~~E=\sqrt{p^2+m_{\rm SM}^2},
\end{equation}
and the fact that at the root $p_*=\sqrt{s}/2$ we have $E_1+E_2=\sqrt{s}$, we get
\begin{equation}
    d\Phi=\frac{d\Omega}{32\pi^2}=\frac{dcd\phi}{32\pi^2},
\end{equation}
where $c=\cos\theta$. Next we choose the z axis aligned with the momentum of the outgoing observed particle $q$ so that $q^\mu=(E,0,0,E)$. Parameterizing the other particle as
$l^\mu=(\omega,\omega \sin\theta_l,0,\omega\cos\theta_l)$ and defining
\begin{equation}
    u=\frac{1-\cos\theta_l}{2},
\end{equation}
we have
\begin{equation}
    d\Pi_l=\frac{\omega d\omega du}{4\pi^2},~~~s=(q+l)^2=4Eu\omega.
\end{equation}
Since all the relevant matrix elements have the form
\begin{equation}
    \vert \mathcal M \vert^2=\frac{s^2}{16 M_{\rm pl}^4}w_iP(\cos\theta_{\rm sc}),
\end{equation}
where $\theta_{\rm sc}$ is the scattering angle and the numerical prefactor is our choice of normalization, we can write the collision term as
\begin{equation}
    C(E,T)=\frac{E}{128\pi^3 M_{\rm pl}^4}w_i\int_0^\infty d\omega \omega^3\int_0^1u^2 du\int_{-1}^1 dc \langle P(\cos\theta_{\rm sc})\rangle_\phi f_1(E_1,T)f_2(E_2,T),
    \label{eq:numInt}
\end{equation}
where 
\begin{equation}
    \langle P(\cos\theta_{\rm sc})\rangle_\phi=\frac{1}{2\pi}\int_0^{2\pi}d\phi P(\cos\theta_{\rm sc}).
\end{equation}
We denote the direction of the momenta $p_1$ and $p_2$ in the center-of-mass frame by $\hat{n}_*$, so that $p_{1,2}=\frac{\sqrt{s}}{2}(1,\pm\hat{n}_*)$. In the thermal frame where the Fermi-Dirac distribution holds the center-of-mass has a nonzero momentum $P^\mu=(P^0,P)$. Boosting to the center-of-mass frame momentum $P^\prime$ we get
\begin{equation}
    P^\prime=\gamma_P(|P|-\beta_PP^0).
\end{equation}
Setting $P^\prime=0$ gives us $\beta_P=\frac{|P|}{P^0}$ and therefore $\gamma_P=(1-\beta^2_P)^{-1/2}=\frac{P_0}{\sqrt{s}}$.
Boosting the center-of-mass frame energies back to the thermal frame we get
\begin{equation}
    E_{1,2}^{\rm thermal}=\frac{E+\omega\pm|P|c}{2},
\end{equation}
where $c=\cos\theta=\hat{n}_*\cdot \hat{P}$ is the angle of the incoming particles in the center-of-mass frame. Here we used
\begin{equation}
    P^\mu=q^\mu+l^\mu,~~~\vert P\vert=\sqrt{(E+\omega)^2-4Eu\omega}.
\end{equation}
Next we define the dimensionless variables
\begin{equation}
    y=\frac{E}{T},~x=\frac{\omega}{T},~S=y+x,~\Delta=y-x,~D=\frac{\vert P \vert}{T}=\sqrt{(y+x)^2-4yxu}.
\end{equation}
Using these the boosted Fermi-Dirac distribution can be written as
\begin{equation}
    f_1(y,x,c)f_2(y,x,c)=\frac{1}{1+e^S+2e^{S/2}\cosh(Dc/2)}.
    \label{eq:FDdistribution}
\end{equation}
Writing the observed particle four momentum in the center-of-mass frame
\begin{equation}
    q_*=\frac{\sqrt{s}}{2}(1,\hat{q}_*)
\end{equation}
and boosting to the thermal frame we get
\begin{equation}
    E=\gamma_P\frac{\sqrt{s}}{2}(1+\beta_Pa),~~~a=\hat{q}_*\cdot \hat{P}.
\end{equation}
In dimensionless variables we get
\begin{equation}
    a=\frac{\Delta}{D}.
\end{equation}
The scattering angle in the center-of-mass frame is $\cos\theta_{SC}=\hat{n}_*\cdot\hat{q}_*$. Parameterizing
\begin{equation}
    \hat{q}_*=(\sin\theta_{q}\cos\phi_q,\sin\theta_{q}\sin\phi_q,a),
\end{equation}
\begin{equation}
    \hat{n}_*=(\sin\theta\cos\phi_n,\sin\theta\sin\phi_n,c),
\end{equation}
we get
\begin{equation}
    \cos\theta_{\rm sc}=ac+\sqrt{1-a^2}\sqrt{1-c^2}\cos\phi,
\end{equation}
where $\phi$ is the difference between $\phi_n$ and $\phi_q$. Further defining
\begin{equation}
    A=ac,~B^2=(1-a^2)(1-c^2)
\end{equation}
we can write $\cos\theta_{\rm sc}=A+B\cos\phi$. In terms of the dimensionless variables we have
\begin{equation}
    A=\frac{(y-x)c}{D},~B^2=\frac{4yx(1-u)(1-c^2)}{D^2}.
\end{equation}
Now $ \langle P(\cos\theta_{\rm sc})\rangle_\phi$ can be easily calculated. We computed the rest of the integral \eqref{eq:numInt} numerically in this paper. 

In the case of vector and scalar bosons, the Fermi-Dirac distribution has to be replaced by the Bose-Einstein one. This amounts to changing the sign of the cosh term in \eqref{eq:FDdistribution}. We can summarize the resulting terms as
\begin{equation}
    F(y,x,c,\eta)=\frac{1}{\eta^2+e^S+2\eta e^{S/2}\cosh(Dc/2)},
\end{equation}
where $\eta=1$ for Fermi-Dirac distribution, $\eta=-1$ for Bose-Einstein distribution and $\eta=0$ for Maxwell-Boltzmann distribution. 

For the total abundance we can compute the integral fully analytically. Start by writing 
\begin{equation}
    S(T)=\int_0^\infty dy\frac{y^3 w_i T^8}{256\pi^5 M_{\rm pl}^4}\int_0^\infty dx x^3\int_0^1u^2 du\int_{-1}^1 dc \langle P(\cos\theta_{\rm sc})\rangle_\phi f_1(E_1,T)f_2(E_2,T).
\end{equation}
Changing the variables to $S, D, a$ instead of $y, x, u$ we get
\begin{equation}
    y^3x^3 u^2 dy dx du=\frac{D^2(S^2-D^2)^2}{64}dS dD da.
\end{equation}
Now there are no $a$ dependent prefactors, neither is there any $a$ dependence in $f_1(E_1,T)f_2(E_2,T)$. The integral
\begin{equation}
    K=\int_{-1}^1 da\langle P(\cos\theta_{\rm sc})\rangle_\phi 
\end{equation}
is taken over the entire range of possible angles of the outgoing particle and therefore cannot depend on $c$. Performing the integration for the various kernels obtained from the stripped matrix elements we get
\begin{align}
    K_{\rm scalar}=\frac{4}{15},~K_{\rm vector}=\frac{8}{5},~K_{\rm fermion}=\frac{8}{5},~K_{\rm EC}=\frac{8}{3},~K_{\rm int}=0.
\end{align}
The interference term integrates to zero exactly as we computed earlier in the paper. We are left with
\begin{equation}
    \frac{K w_i T^8}{16384\pi^5 M_{\rm pl}^4}\int_0^\infty dS\int_0^S dD D^2(S^2-D^2)^2\int_{-1}^1dc F(S, D, c).
\end{equation}
Transferring to a new variable $v=Dc$, changing the order of integration between $D$ and $v$ and calculating the integral in $D$, we get
\begin{equation}
    \frac{K w_i T^8}{98304\pi^5 M_{\rm pl}^4}\int_0^\infty dS\int_{-S}^S dv(S^2-v^2)^3 F(S, v).
\end{equation}
Finally, taking the variables $v_+=(S+v)/2$ and $v_-=(S-v)/2$ the integral factorizes into a product of two identical ones
\begin{equation}
    S=\frac{K w_i T^8}{768\pi^5 M_{\rm pl}^4}\left[\int_0^\infty dp p^3 f(p)\right]^2,
\end{equation}
where $f(p)$ is the Fermi-Dirac or Bose-Einstein distribution. The integral evaluates to 
\begin{equation}
    \int_0^\infty dp p^3 f(p)=\frac{7\pi^4}{120}
\end{equation}
for the Fermi-Dirac case and
\begin{equation}
    \int_0^\infty dp p^3 f(p)=\frac{\pi^4}{15}
\end{equation}
for the Bose-Einstein one. 
\bibliography{ref}

\end{document}